\documentclass[12pt]{article}

\usepackage{geometry}
\usepackage[utf8]{inputenc}
\usepackage{adjustbox}
\usepackage{amsmath}
\usepackage{url}
\usepackage[square]{natbib}
\usepackage{booktabs, caption, makecell}

\title{\Large Robust model-based estimation for binary outcomes in genomics studies}
\author{\large Suyoung Park$^1$, Alexander E. Lipka$^2$, Daniel J. Eck$^1$\\[1em]
\small 1. Department of Statistics, University of Illinois Urbana-Champaign \\
\small 2. Department of Crop Sciences, University of Illinois Urbana-Champaign}
\date{\normalsize October, 2021}

\begin{document}
\maketitle
\begin{abstract}

In quantitative genetics, statistical modeling techniques are used to facilitate advances in the understanding of which genes underlie agronomically important traits and have enabled the use of genome-wide markers to accelerate genetic gain. The logistic regression model is a statistically optimal approach for quantitative genetics analysis of binary traits. To encourage more widespread use of the logistic model in such analyses, efforts need to be made to address separation, which occurs whenever a specific combination of predictors can perfectly predict the value of a binary trait. Data separation is especially prevalent in applications where the number of predictors is near the sample size. In this study we motivate a logistic model that is robust to separation, and we develop a novel prediction procedure for this robust model that is appropriate when separation exists. We show that this robust model offers superior inferences and comparable predictions to existing approaches while remaining true to the logistic model. This is an improvement to previously existing approaches which treats separation as a modeling shortcoming and not an antagonistic data configuration. Previous approaches, therefore, change the modeling paradigm to consider separation that, before our robust model exists, is problematic to logistic models. Our comparisons are conducted on several didactic examples and a genomics study on the kernel color in maize. The ensuing analyses reaffirm the billed superior inferences and comparable predictive performance of our robust model. Therefore, our approach provides scientists with an appropriate statistical modeling framework for analyses involving agronomically important binary traits.

\end{abstract}

\section{Introduction}

The application of quantitative genetics approaches to crops has facilitated advances in the basic understanding of which genes underlie agronomically important traits, and has enabled the use of genome-wide markers to accelerate genetic gain. For example, the use of statistical models in genome-wide association studies has provided insight into the role of pleiotropy in the genetic architecture of leaf and inflorescence-related traits in maize \citep{rice2020multi}. Similarly, multi-kernel genomic prediction (GP) models that include environmental covariate information have made it possible to accurately predict genomic estimated breeding values (GEBVs) for grain yield in wheat in specific environments \citep{jarquin2014reaction}. Although less commonplace than quantitative traits, agronomically important binary traits do occur. For example the color of kernels in maize is often dichotomized into a binary trait (white versus yellow; \citep{Romay2013}), and breeding for kernel color is a critical step for increasing bioavailablity of provitamin A carotenoids in maize grain \citep{chandler2013genetic, harjes2008natural}. Thus, the  application of quantitative genetic analysis to binary traits has great potential to have a meaningful impact on future agronomic efforts. However, a major setback is that some of the most widely-used quantitative genetics approaches in agronomy do not account for the dichotomous configuration of a binary trait. Consequently, direct application of state-of-the-art quantitative genetic approaches to study binary traits could result in negative statistical ramifications, including inadequate control of inflation of test statistics due to subpopulation structure (as shown in, e.g. \citep{shenstone2018assessment}).

From a statistical perspective, the most logical model to apply to the analysis of binary traits is the logistic regression model \citep{agresti_categorical_2013}, in which a binary outcome variable depends on a set of explanatory variables. However, a major setback of the logistic regression model is that it breaks down whenever there exists a linear combination of the explanatory variables in which a binary outcome of "0" or "1" is guaranteed to be observed. Whenever this phenomenon, known as separation, is present in data, the logistic model coefficient estimates are not finite (or unstable), and it becomes impossible to obtain a reasonable estimate of the sizes of marker effects. Given that statistical software typically fails to diagnose this issue  \citep{Eck2021} and that the number of markers included in a logistic regression model is likely to be large and can exceed the sample size (especially in genomic prediction applications), it is critical that the issue of separation in quantitative genetic data is carefully assessed and the potential of remedial statistical approaches are considered. 

The easiest way to deal with  separation, when it is detected, is to remove markers with uninterpretable effect sizes from the model. However, this na\"ive approach often leads us to get rid of the markers with the strongest associations with the binary trait  \citep{Zorn2005}. To eliminate the need to take such drastic measures, modifications to logistic regression have been proposed. For example \cite{Heinze2002} used the Firth's penalized maximum likelihood estimation to reduce the bias of maximum likelihood estimator, and thus obtained reasonable, finite  estimates of marker effects. \cite{Kosmidis2009} generalized this method for the nonlinear exponential family, thus making Firth's penalized likelihood available to traits following any distribution of the exponential family. Additionally, many have proposed a Bayesian framework to handle the estimation problems that arise from separation \citep{Heinze2002, Dunson_2006, Genkin_2007, Gelman2008}. In fact, \citeauthor{Heinze2002}'s method can be interpreted from a Bayesian perspective as the application of Jeffrey's invariant prior. Further examples of approaches used to deal with separation include \cite{Dunson_2006}, which used the mixture prior distributions for the logistic model with large number of markers,  \cite{Genkin_2007} considered the Laplace prior distribution, and \cite{Gelman2008} suggested the Cauchy distribution with center 0 and scale 2.5 as the default choice. This latter approach showed faster and comparable predictive performance in terms of cross-validation predictive error than other methods.

Both bias reduction and Bayesian approaches handled the separation issue by switching the modeling paradigm to accommodate problematic data rather than solving the issue within the original model. \cite{Geyer2009} developed methodology for directly finding maximum likelihood estimates (MLE) when problematic separation exists and the traditional MLE calculations do not converge. This method required a massive computation cost which makes it time consuming to apply in practice. \cite{Eck2021} proposed new, faster and scalable methodology to find the MLE in the completion when MLE does not exist. The methods of \cite{Geyer2009} and \cite{Eck2021} work by first detecting separation when it exists, and then providing one-sided confidence intervals for problematic data points that exist in the separated space and usual two-sided confidence intervals for non-problematic data points. \citeauthor{Eck2021}'s method is implemented in the R package \texttt{glmdr}, software which detects and remedies separation in logistic regression. 

In this study we developed a prediction framework compatible with \cite{Eck2021}'s method. This prediction method did not previously exist. We begin by motivating the logistic regression model for quantitative genetics analysis of binary, that is, Bernoulli-distributed quantitative traits. We then describe the problem of separation in the data, and compare different techniques (Bayesian, penalized likelihood, and maximum likelihood estimation) for handling separation on several datasets, as well as a quantitative genetics dataset from maize \citep{Romay2013}. Next we assess performance of these techniques on their inferential and predictive ability. Because the MLE asymptotically achieves the Cram{\'e}r-Rao lower bound, we expected the MLE technique in \cite{Eck2021} to yield the tightest inferences among all techniques under consideration. This finding is confirmed in all datasets that we considered. We then motivate our novel prediction procedure within the methodological context of \cite{Eck2021}. We expected that our developed method and the other considered methods to exhibit even predictive performance with a computational edge towards the Bayesian techniques that we considered. While the method of \cite{Eck2021} is far more computationally convenient than that of \cite{Geyer2009} it is still rather involved when adapted for prediction. Ultimately we want to develop methodology in \cite{Eck2021} that is applicable for genomic prediction when there may be far more predictors than cases. The prediction procedures developed here are an important step in that direction.

\section{Materials and Methods}

\subsection{Logistic Regression}
The logistic regression is a special case of the generalized linear model which the outcome variable follows a Bernoulli distribution (i.e., $y \in \{0,1\}$) \citep{GLM_paper}. By convention, we encode 1 as a ``success" and 0 as a ``failure." In logistic regression the conditional success probability at a particular $x$ is modeled as
\begin{equation}\label{logistic_model}
\text{Pr}(Y = 1 | X = x) = \frac{\exp{(x^{T}\beta)}}{1+\exp{(x^{T}\beta)}} = \pi(x),
\end{equation}
where $\beta$ is an unknown canonical parameter vector (coefficient vector), $X$ and $Y$ are the covariate and outcome random variables, and $x$ is an observed value.

From the linear regression's point of view, this logistic regression is equivalent to:
\begin{equation}\label{link_function}
g(\pi(x)) = \log\left(\frac{\pi(x)}{1-\pi(x)}\right) = x^T\beta
\end{equation}
where $g(x) = \log(\frac{x}{1-x})$ is a logit link (log-odds ratio).

Therefore, as in classical ordinary least squares (OLS) regression, we can estimate model parameters using maximum likelihood estimation. Statistical inferences about model parameters can be obtained from estimates of the Fisher information. Unlike in OLS regression, estimates for $\hat\beta$ are not given in closed form. The log-likelihood function for the logistic regression model is
\begin{equation}\label{likelihood}
\log L(\beta|Y) = \sum_{i=1}^{n} y_i\log{(\pi(x_{i}))} + (1-y_i)\log{(1-\pi(x_{i}))},
\end{equation}
one then obtains $\hat\beta$ by solving the score function equation
\begin{equation}\label{logit_Score}
\frac{\partial \log{L(\beta|Y)}}{\partial \beta} = 
\sum_{i=1}^{N} (y_i - \log{(\pi(x_{i}))}) x_i^T =
\sum_{i=1}^{N} [y_i + \log{(1+\exp{(-x_i^T\beta)})}] = 0.
\end{equation}
Conventional softwares finds $\hat\beta$ through Fisher-scoring or iteratively reweighted least squares algorithms \citep[Chapter 4]{agresti_categorical_2013}. We then obtain inferences using an estimate of the Fisher information matrix evaluated at the MLE solution $\hat{\beta}$
\begin{equation}\label{logit_Fisher}
\widehat{\text{Var}(\beta)} = [I(\hat\beta)]^{-1} = \left(-E\left[\frac{\partial^2 \log{L(\beta|Y)}}{\partial \beta_i \partial \beta_j}\right]\right)^{-1}\Big\vert_{\beta = \hat\beta}.
\end{equation}
Conventional software provides \eqref{logit_Fisher}.

\subsection{Complete Separation}
\label{sec:complete_separation}

\begin{figure}[!ht]
  \centering
    \includegraphics[width=0.5\textwidth]{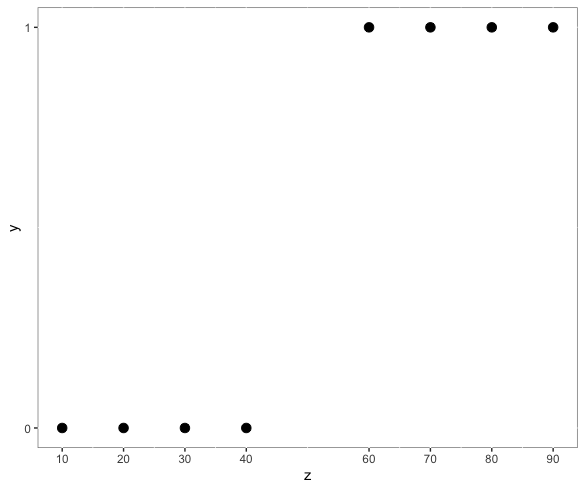}
    \caption{Example of complete separation from Section 6.5.1 of \cite{agresti_categorical_2013}. The conventional MLE of a logistic model does not exist.}\label{Fig:Complete_Separation_Example}
\end{figure}

Traditional maximum likelihood estimation for logistic regression does not work well when there is complete or quasi-complete separation in the data, a problem that is widespread in applications \citep{Geyer2009}. \cite{agresti_categorical_2013} defined complete separation when there exists a vector $b$ such that 

\begin{equation}\label{definition_complete_separation}
\begin{split}
x_{i}^{T} b &> 0 \text{ whenever } y_i = 1, \\
x_{i}^{T} b &< 0 \text{ whenever } y_i = 0.
\end{split}
\end{equation}
That is, complete separation occurs when the one or more covariates can perfectly predict the outcome variable \citep{AlbertAnderson1984}. For example, as shown in the Figure~\ref{Fig:Complete_Separation_Example}, consider the following case that when $z$ is less than 50, all corresponding $y$ are $0$ and when $z$ is greater than 50, all corresponding $y$ are $1$. Suppose we are interested in a simple logistic regression model $x_i^T = [1, z_i]$. Then this data is completely separated with $b = [-50, 1]^T$. Moreover, we have $\hat{\pi}(x) = 0$ for $z < 50$ and $\hat{\pi}(x) = 1$ for $z > 50$.

\begin{figure}[!ht]
  \centering
    \includegraphics[width=\textwidth]{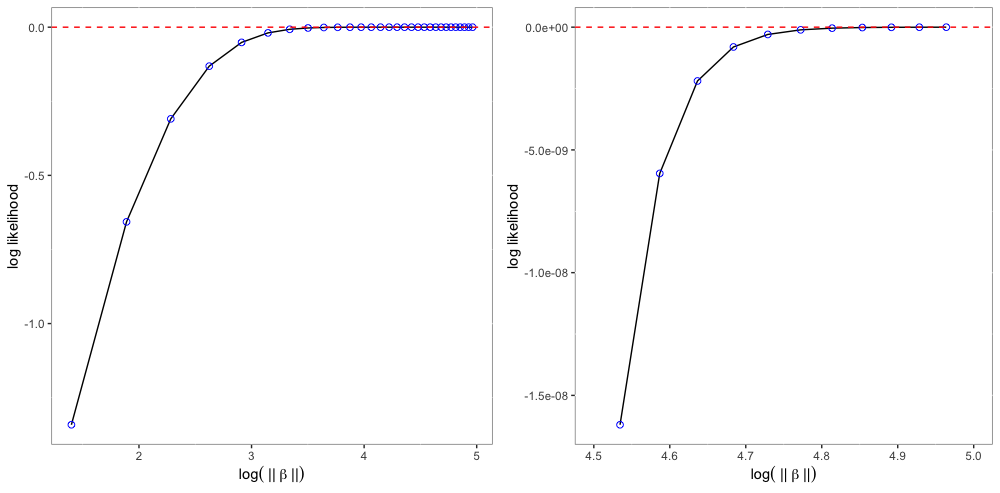}
    \caption{\textbf{Left panel}: Log likelihood values of logistic model at different working estimates. Blue dot represents the log likelihood value at each iteration. \textbf{Right panel:} Zoom in view of a log likelihood values of logistic model where log of norm of working estimates lie between 4.5 and 5.}\label{Fig:asymptotes}
\end{figure}

The parameter estimates $\hat\beta$ are ``at infinity'' when separation exists in the data. The iteration based estimation algorithms under the hood of conventional software packages provide a sequence of estimates that goes to infinity, and the log likelihood becomes flat when evaluated along this sequence. The left panel of Figure \ref{Fig:asymptotes} shows the log likelihood of logistic model for this example with different working estimate from $\texttt{glm}$ function in R. We can see that each iteration, the norm of $\beta$ becomes larger and asymptote of the log likelihood value goes to infinity. The right panel of Figure \ref{Fig:asymptotes} is the zoomed part of the left panel of Figure \ref{Fig:asymptotes} where the log of norm of working estimates is between 4.5 and 5. It displays the log likelihood value still approaches near zero although the left panel of Figure \ref{Fig:asymptotes} looks flat in the same region. 

Standard statistical inferences are not appropriate when separation exists. The standard errors of predicted probabilities of success are very small, which leads to extremely narrow confidence intervals for each observation. Unfortunately, none of common statistical software such as R, SAS and Python can handle the separation issue properly and uninformed users sometimes uses the wrong model without knowing it \citep{R_software, SAS_software, Python_software}. The \texttt{glmdr} software package \citep{glmdr_package} is designed to provide users with a description of the separation problem, and provide statistical inferences when it occurs. This package did not have a prediction routine before this work.

Quasi-complete separation is another case of separation that there are both a success and a failure on the hyperplane that separates the successes from the failures \citep{lesaffre_albert_1989}. For instance, we can consider additional two points that $z = 50$ with $y=1$ and $y=0$ to the previous complete separation example. That is, we have $y_i = 0$ for $z \leq 50$ and $y_i =1$ for $z \geq 50$. In this case, the maximized log likelihood is always negative and we experience same phenomenon as the complete separation case.

\subsection{Mean-value Parameters}

The mean-value parameter is the conditional expectation of the response variable given a covariate value $\text{E}(Y|X=x)$. In logistic regression the mean-value parameter is modeled as \eqref{logistic_model} where $\text{E}(Y|X=x) = \text{Pr}(Y=1|X=x)$. The mean-value parameter is often the most important parameter for inference, we want to know the success probability at a particular $x$. An example would be to model the probability of the kernel color being white or yellow given the marker genotypes encoded in the covariate vector $x$. Unlike in linear regression, this parameter is not easily interpreted from $\beta$. Thus inferences about $\beta$ does not provide useful inferences for $\text{Pr}(Y=1|X=x)$. Furthermore, the natural constraints on a conditional probability corresponding to a binary outcome variable require an alteration to the linear model. We expand on these points in the next paragraph.

In linear regression we can easily obtain $\pi(x) = \text{E}(Y|X=x)$ from $\beta$ since $\text{E}(Y|X=x) = x^{T}\beta$. Plugging in $\hat{\beta}$ produces the MLE for this expectation $\hat{\pi}(x) = \widehat{\text{E}}(Y|X=x) = x^{T}\hat\beta$ with $x$ fixed. On the other hand, in the logistic model, $\pi(x) = \text{E}(Y|X=x) = \text{Pr}(Y = 1 | X=x)$ where $\log(\frac{\pi(x)}{1-\pi(x)}) = x^T\beta$. Thus, $\beta$ does not offer an easy interpretation about changes in the expected outcome as the covariates change, and it is therefore less useful as a parameter for understanding how $\pi(x)$ changes with $x$. The mean-value parametrization is the primary parameter of interest in both regression contexts, but in linear regression the mean-value parameter and $\beta$ are interchangeable.

Another benefit of the mean-value parameterization over $\beta$ in the logistic regression model is when complete separation exists. When complete separation exists $\beta$ is estimated to be at infinity while $\pi(x)$ is estimated to be 0 or 1. We discuss complete separation and methods which address it in the next Section.

\subsection{One-Sided Confidence Interval}

Standard statistical inferences cannot be performed in the presence of separation. In this setting, statistical inferences will be made for estimated mean-value parameters. Separation necessitates the use of one-sided confidence intervals for estimated mean-value parameters since $\beta$ is estimated to be at infinity. This is because when separation exists $\pi(x)$ is estimated to be 0 or 1, the boundary of allowable values, and two-sided intervals would therefore include values that are lower than 0 or greater than 1. The original concept for the one-sided intervals that we use in this paper can be found in Section 3.16 of \cite{Geyer2009} with implementation details can be found in Section 4.3 of \cite{Eck2021}. These one-sided intervals are specifically tailored to handle separation and they are what form our robust logistic regression model.

\begin{figure}[!ht]
  \centering
    \includegraphics[width=0.5\textwidth]{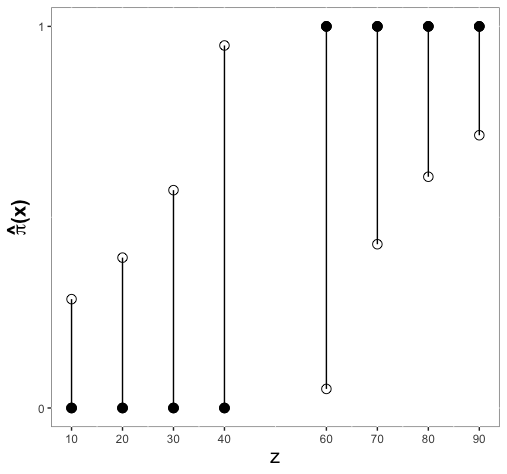}
    \caption{One-sided 95\% confidence interval for the example of complete separation from Section \ref{sec:complete_separation}. Solid dot represents the observed value and bar shows the interval. $\hat{\pi}(x)$ is the estimated probability of a success given $z$ where $x = (1,\; z)^T$.}\label{Fig:One_sided_CI}
\end{figure}

For completeness we briefly explain how we construct these one-sided confidence interval for mean-value parameters. One endpoint of the one-sided interval is constrained to be the observed outcome variable (i.e., lower bound if $y_i=0$ and upper bound if $y_i=1$), and the other endpoint is obtained by solving the optimization problem:

\begin{equation} \label{CI_logistic}
\begin{split}
    \text{minimize } & \quad -\theta_k \\
    \text{subject to } & \sum_{i \in I}[y_{i} \log{(\pi(x_{i}))} + (1 - y_i) \log{(1-\pi(x_{i})})] - \log{(\alpha) \geq 0},
\end{split}
\end{equation}
where $\theta_k = x_k^T\beta$ for any $k \in I$, $\textit{I}$ is a index of problematic points that cause the separation, $\pi$ is a mean-value parameter, and $\alpha$ is a significance level. For example, Figure \ref{Fig:One_sided_CI} shows the one-sided confidence interval for the complete separation example we discussed in Section \ref{sec:complete_separation}. We can see the confidence interval increases as $z$ increases until $z = 40$ then it starts to decrease as $z$ increases from $z = 60$. Also, we have a widest interval where $z = 40$ and $z = 60$ with the length of intervals, $1-\alpha$. It means our uncertainty on estimation keep increases from $z=10$ to $z=40$ and we have the highest uncertainty near the separation occurs. Then it diminishes as it furthers away from the boundary of the separation. In $\texttt{glmdr}$, $\texttt{inference}$ function provides this confidence intervals using the sequential quadratic programming (SQP) to solve the constrained nonlinear problem \eqref{CI_logistic}. 

\subsection{Prediction}

Model based prediction is different when data separation is present. Standard techniques fail in the presence of data separation. In the absence of separation we can compute the predicted value for new data point from the logistic model using $\hat{\pi}(x_{\text{new}}) = (1 + \exp{(-x_{\text{new}}^T \hat{\beta})})^{-1}.$ However, when complete separation is present, this approach breaks down since $\hat{\beta}$ is at infinity in this setting, and this implies that $\hat{\pi}(x_{\text{new}})$ is either 0 or 1. Standard estimates of variability suffer from a similar problem. Another difficulty is due to uncertainty in the separation itself. Data separation occurs with probability tending to zero as the sample size increases with the number of predictors fixed. Data separation is therefore a sampling issue, not a modeling issue. We propose a new method for prediction that addresses the practical and conceptual difficulties of prediction when separation exists. 

This method is as follows: 
\begin{enumerate}
	\item Pick a point $x_{\text{new}}$ for which a prediction is desired and there exists separation at $x_{\text{new}}$. A prediction at $x_{\text{new}}$ is either 0 or 1 using traditional methods.
	\item We then combine this point with the observed data. We will make a model-based estimate at $x_{\text{new}}$ by fitting separate logistic regression models, one outcome label $y_{\text{new}}=0$ and the other with $y_{\text{new}}=1$. Fitting two separate models in this way is intended to address the uncertainty in the data separation.
	\item  We then compute the estimated probability of a success for new data points, $\hat{\pi}(x_{\text{new,0}})$ and $\hat{\pi}(x_{\text{new,1}})$. Note that one of $\hat{\pi}(x_{\text{new,0}})$ or $\hat{\pi}(x_{\text{new,1}})$ will be 0 or 1 and the other will not be, this is because $y_{\text{new}}=0$ or $y_{\text{new}}=1$ decreases the uncertainty in the separation by adding a pseudo-outcome in its favor, while the other pseudo-outcome alleviates the separation at $x_{\text{new}}$.
	\item  We now combine $\hat{\pi}(x_{\text{new,0}})$ and $\hat{\pi}(x_{\text{new,1}})$ to form a prediction using model averaging. Our model averaging procedure judges the fit of each model based on weights similar to the the Akaike weights in \cite{burnham_anderson_2002}. These weights are 
$$
  w_j = \frac{{\exp(-\frac{IC_j}{2})}}{\exp(-\frac{IC_0}{2})+\exp(-\frac{IC_1}{2})},
$$ 
  where $IC_j$ is the information criteria of model $j$.
  \item We then can calculate the model averaged estimate, $\hat{\pi}^{*}(x_{\text{new}}) = \sum_{j=0}^{1} w_j \hat{\pi}(x_{\text{new,j}})$. We used Akaike information criteria corrected (AICc) as $IC_j$. The primary reason for its use is that AICc does not have an overfit problem when the sample size is small \citep{AICc_Paper}. The presence of data separation is an indication that one is not close to asymptopia, the sample size is small in this sense.
  \item  We then label $y^{*}_{\text{new}}$ as $1$ if $\hat{\pi}^{*}(x_{\text{new}}) \geq C^*$ and $0$ if $\hat{\pi}^{*}(x_{\text{new}}) < C^*$ where $C^*$ is the optimal cut-off that maximizes the overall accuracy. The main motivation of using optimal cut-off is that threshold of $0.5$ produces unreliable and poor model accuracy when the outcome variable is highly unbalanced \citep{Freeman_2008}.
\end{enumerate}
These six steps are outlined in Figure \ref{Fig:FlowChart}. Detailed implementation of this prediction method and examples are given in the Supplementary Materials.

\begin{figure}[!ht]
  \centering
    \includegraphics[width=1\textwidth]{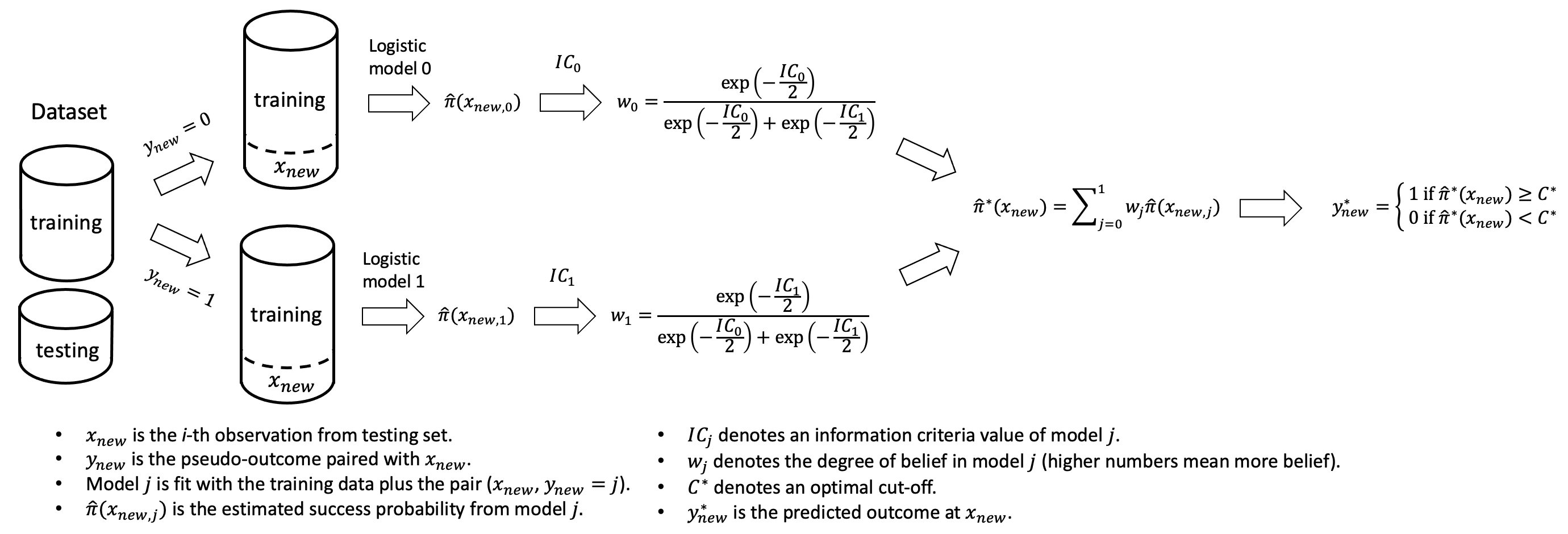}
    \caption{Flowchart for the prediction method.}
    \label{Fig:FlowChart}
\end{figure}

We also calculate the prediction intervals for model averaged estimate, $\hat{\pi}^{*}(x_{\text{new}})$. We construct the Wilson intervals [\citeyear{wilson_1927}] for predicted probabilities. Wilson intervals show better coverage probability although $\hat{\pi}^{*}(x_{\text{new}})$ is near $0$ and $1$ boundaries in comparison to the standard binomial confidence interval because Wilson intervals are asymmetric \citep{Brown2001}. 

\subsection{Model Performance}

\subsubsection{Competing models}

We compare the performance of direct maximum likelihood estimation fit with \texttt{glmdr} with Bayesian generalized linear regression models fit with \texttt{bayesglm}, bias reduction generalized linear regression models fit with \texttt{brglm}, and the classical linear model fit with \texttt{lm}. Briefly, these different models are summarized as:
\begin{enumerate}
\item \texttt{glmdr}: \texttt{glmdr} stands for ``generalized linear models done right" \citep{glmdr_package}. It solves the separation issue within the original model settings. \texttt{glmdr} provides the inference by estimating the probability of success based on MLE in Barndorff-Nielsen completion [\citeyear{barndorff-nielsen_1978}], and calculating one-sided confidence intervals. \texttt{glmdr} makes a prediction based on the model averaged estimate.
\item \texttt{bayesglm}: \texttt{bayesglm} is the generalized linear model with student-t prior distribution \citep{Gelman2008}. It scales the original data then uses Cauchy distribution as a prior distribution on the coefficients to handle the separation issue. 
\item \texttt{brglm}: \texttt{brglm} denotes ``bias reduction in generalized linear model" \citep{Kosmidis2009}. It modifies the score function with powers of the Jeffreys prior as penalty to produce the finite coefficients (penalized MLE) to deal with the separation issue.
\item \texttt{lm}: \texttt{lm} stands for ``linear model" that provides the estimates based on ordinary least squares method. It is one of the most common statistical models for regression problems.
\end{enumerate}

\subsubsection{Performance metrics}

We assess model performance using several metrics, these are:
\begin{enumerate}
	\item In-sample estimation accuracy in which we count the number of correctly classified observations in the training set divided by total number of observations in the training set.
	\item Out-of-sample prediction accuracy which is assessed via leave-one-out cross validation (LOOCV). Since the predicted value of linear model does not have to fall into $[0,1]$ range, we assign $1$ for any predicted values greater than $1$ and $0$ for negative values.
	\item The average length of each confidence and prediction interval.
	\item The computational cost in seconds of each prediction method. This cost is calculated using the \texttt{proc.time} function in R.
\end{enumerate}

\subsection{Data}

We provide inference and prediction results for the maize data, a dataset corresponding to an endometrial cancer study, and several didactic examples. These didactic examples are motivated in \cite{agresti_categorical_2013} and \cite{Geyer2009}, they are designed to be understandable for audiences new to separation in the data. A description of these data sets follows: \\

\noindent\textbf{Complete separation}: This example comes from  \cite{agresti_categorical_2013} and is discussed in Section~\ref{sec:complete_separation}. In this example, there is a binary outcome variable, $y \in \{0,1\}$ and one covariate variable, $z$, with 8 data points. Specifically, $y_i = 1$ at $z = 10, 20, 30, 40, $ and $y_i = 0$ at $z = 60, 70, 80, 90$. Since $y$ could be completely separable by $z$, we observed the complete separation in this example. \\

\noindent\textbf{Quasi-complete separation}: 
 This example is an extension of the complete separation example in \cite{agresti_categorical_2013} with two points added, $y_i=1$ and $y_i=0$ at $z = 50$. This is an example of quasi-complete separation. \\

\noindent\textbf{Quadratic logistic regression model}: This example comes from Section 2.2 of Geyer [\citeyear{Geyer2009}]. There is one binary outcome variable $y \in \{0,1\}$ and one covariate variable $z$ which takes integer values from 1 to 30. The outcome variable was $y_i =1$ when $ 12 < z_i < 24 $ and $y_i = 0$ otherwise. A quadratic logistic model is considered in this example and complete separation is observed. \\

\noindent\textbf{Endometrial Cancer Study}: This example comes from \cite{Heinze2002}. 
The main purpose of this study was to describe histology of cases (HG) in terms of three risk factors: neovasculation (NV), endometrium height (EH) and pulsatility index of arteria uterina (PI). The outcome variable had 30 patients classified grading 0-II for histology (HG = 1) and 49 patients for grading III-IV (HG = 0). There were 13 patients who had neovasculization (NV = 1) and absent for 66 patients (NV = 0). Pulsatility index (PI) ranges from 0 to 49 with mean of 17.38 and median of 16.00, and endometirum height (EH) ranges from 0.27 to 3.61 with mean of 1.662 and median of 1.640. Quasi-complete separation was observed in this example, this separation is determined by NV. \\ 

\noindent\textbf{Maize data}: This example comes from \cite{Romay2013}, and it consists of 2,815 maize lines. The binary outcome variable is the kernel color, where 1 indicated non-white kernel color and 0 indicated white kernel color. We fitted a logistic regression model with kernel color as the outcome variable and covariate variables consisting of subpopulations and 24 DNA markers surrounding the \textit{psy1} gene \citep{palaisa2004long, chandler2013genetic, ford2000inheritance}. The \textit{psy1} gene has been shown to be each marker had a value from 0 to 1, where 0 and 1 indicated homozygous genotypes and a value in (0,1) indicated that missing genotypic data were imputed. In the final data set, 309 lines had a white kernel and 1,238 had non-white kernel color. These maize lines were subdivided into six subpopulations, namely 115 non-stiff stalk, 54 popcorn, 120 stiff stalk, 116 sweet corn, 159 tropical, and 983 unclassified. In this example, there was no separation issues when we used a single marker for covariate. However, we had a separation issue for saturated model with 24 DNA markers and subpopulations.

\subsection{Software used}

We implemented our methodology in R package $\texttt{glmdr}$. We used R version 3.6.1 and the required R packages for $\texttt{glmdr}$ is $\texttt{nloptr}$ version 1.2.2.2. To compare its performance, we considered $\texttt{arm}$ version 1.11-1, $\texttt{brglm2}$ 0.7.0, and $\texttt{stats}$ version 3.6.1. To determine the optimal cut-off for the logistic regression, we used $\texttt{PresenceAbsence}$ version 1.1.9. For visualization, data wrangling and experiments, we used $\texttt{ggplot2}$ version 3.3.3, $\texttt{gridExtra}$ version 2.3, $\texttt{latex2exp}$ version 0.4.0, $\texttt{foreach}$ version 1.4.7, $\texttt{doParallel}$ version 1.0.15, and $\texttt{tidyverse}$  version 1.2.1. Further details are included in the technical reports. $\texttt{glmdr}$ is available on \url{https://github.com/DEck13/complete_separation}.

\section{Results}

\subsection{Inference}
\label{sec:inference}

Inferences are for mean-value parameters $\pi(x)$ for observed values of $x$. We report the in-sample accuracy for all observations and confidence intervals for problematic observations that raise the (quasi) complete separation issue to compare each method. For confidence intervals, we compute the average length of one-sided confidence interval for $\texttt{glmdr}$ and average length of Wilson intervals for $\texttt{bayesglm}$,  $\texttt{brglm}$, and linear models. In Table \ref{Tab:inference}, we can see all methods show the equivalent in-sample accuracy for the complete separation and quasi-complete separation examples. Meanwhile, the logistic models, $\texttt{glmdr}$, $\texttt{bayesglm}$, and  $\texttt{brglm}$, display the higher in-sample accuracy for quadratic, endometrial, and maize examples in comparison to the linear model. Within these examples, $\texttt{glmdr}$ has the highest in-sample accuracy in maize example than other two logistic models. For confidence intervals, $\texttt{glmdr}$ demonstrates the smallest length in all examples. Especially, in quadratic and endometrial examples, its lengths of confidence intervals are significantly smaller than other methods. Two logistic models, $\texttt{bayesglm}$ and $\texttt{brglm}$ generally shows smaller lengths of confidence intervals but they are not highly different from that of linear model in all examples. This result suggests that linear model perform worse than logistic model remedies, and $\texttt{glmdr}$, which solves the complete separation within the MLE framework, produces the most accurate inferences.

\begin{table}[!ht]
\caption{Model performances for all examples. Estimation accuracy and length of confidence intervals are displayed.}
\caption*{\footnotesize \textit{glmdr denotes Generalized Linear Model Done Right \citep{glmdr_package}, bayesglm denotes Generalized Linear Model with Student-t prior distribution \citep{Gelman2008}, brglm denotes Bias Reduction in Generalized Linear Models \citep{Kosmidis2009}, and linear denotes the multiple linear model using ordinary least squares.}}\label{Tab:inference}
\centering
    \resizebox{\columnwidth}{!}{
    \begin{tabular}{ll|ccccc}
 &  & Complete Separation & Quasi Separation & Quadratic & Endometrial & Maize \\ \hline
accuracy & glmdr & 100 \% & 90 \% & 100 \% & 88.61 \% & 87.14 \% \\
 & bayesglm & 100 \% & 90 \% & 100 \% & 88.61 \% & 87.07 \% \\
 & brglm & 100 \% & 90 \% & 100 \% & 88.61 \% & 87.01 \% \\
 & linear & 100 \% & 90 \% & 90 \% & 86.08 \% & 86.81 \% \\
length & glmdr & 0.550 & 0.308 & 0.199 & 0.194 & 0.563 \\
 & bayesglm & 0.828 & 0.827 & 0.823 & 0.804 & 0.814 \\
 & brglm & 0.835 & 0.831 & 0.811 & 0.808 & 0.826 \\
 & linear & 0.829 & 0.829 & 0.859 & 0.806 & 0.838
\end{tabular}
    }
\end{table}

\subsection{Prediction}
\label{sec:prediction}

To compare the performance of prediction, we compare out-of-sample accuracy, prediction intervals, and computational cost. In Table \ref{Tab:prediction}, we can see all methods show comparable accuracy for the complete separation and quasi-complete separation examples. $\texttt{glmdr}$ shows the highest out-of-sample accuracy in endometrial example where other three methods perform the same. In the quadratic example, $\texttt{brglm}$ performs the best followed by other two logistic models and linear model, but linear model is better than the logistic models in maize example although their differences are not large. This result is surprising because the linear model is generally not recommended for binary classification, yet it shows a better performance than the logistic models. For prediction intervals, overall there is no significant difference between each method. We notice that $\texttt{glmdr}$ has the smallest lengths of prediction intervals in three examples. However, $\texttt{bayesglm}$ produces the smallest interval for the quasi-complete separation, and the linear model produces the smallest interval for the endometrial example.

\begin{table}[!hb]
\caption{Prediction results. length of prediction intervals, and computational cost for all examples.}
\caption*{\footnotesize \textit{glmdr denotes Generalized Linear Model Done Right \citep{glmdr_package}, bayesglm denotes Generalized Linear Model with Student-t prior distribution \citep{Gelman2008}, brglm denotes Bias Reduction in Generalized Linear Models \citep{Kosmidis2009}, and linear denotes the multiple linear model using ordinary least squares.}}\label{Tab:prediction}
\centering
    \resizebox{\columnwidth}{!}{
    \begin{tabular}{ll|ccccc}
 &  & Complete Separation & Quasi Separation & Quadratic & Endometrial & Maize \\ \hline
accuracy & glmdr & 100 \% & 80 \% & 93.33 \% & 87.34 \% & 86.23 \% \\
 & bayesglm & 100 \% & 80 \% & 93.33 \% & 86.08 \% & 86.36 \% \\
 & brglm & 100 \% & 80 \% & 100 \% & 86.08 \% & 86.36 \% \\
 & linear & 100 \% & 80 \% & 90 \% & 86.08 \% & 86.55 \% \\
length & glmdr & 0.822 & 0.859 & 0.807 & 0.848 & 0.837 \\
 & bayesglm & 0.839 & 0.845 & 0.828 & 0.843 & 0.837 \\
 & brglm & 0.843 & 0.847 & 0.813 & 0.844 & 0.837 \\
 & linear & 0.833 & 0.844 & 0.861 & 0.851 & 0.839 \\
cost & glmdr & 0.13 secs & 0.27 secs & 0.31 secs & 1.06 secs & 3.70 mins \\
 & bayesglm & 0.11 secs & 0.12 secs & 0.35 secs & 0.31 secs & 45.35 secs \\
 & brglm & 0.19 secs & 0.19 secs & 0.44 secs & 0.49 secs & 3.80 mins \\
 & linear & 0.07 secs & 0.06 secs & 0.09 secs & 0.14 secs & 4.63 secs
\end{tabular}}
\end{table}

We present the computational cost of each method in Table \ref{Tab:prediction}. In all examples, linear model is much faster than logistic model remedies. Although there is no significant difference in complete separation, quasi-complete separation, quadratic, and endometrial examples, computational cost of $\texttt{glmdr}$ increases in maize example. This increase is due to the execution time for $\texttt{glmdr}$ to solve its internal optimization problem when the data point to be predicted yields separation. Similarly, $\texttt{brglm}$ is as slow as $\texttt{glmdr}$ because it needs to handle the optimization problem to find the penalized MLE for each iteration. However, $\texttt{bayesglm}$ does not suffer this issue because it does not carry the computation for the optimization problem in their method.

Considering all aspects, all of four methods demonstrate comparable out-of-sample accuracy and length of prediction intervals. However, there are several notable differences. $\texttt{glmdr}$ provides the smallest lengths of prediction intervals except in the quasi-complete separation example. It also shows better performance in the endometrial example. But, it may not be scalable to the large datasets due to relatively high computational cost. $\texttt{bayesglm}$ performs well on all examples with the lowest computational cost, which indicates the $\texttt{bayesglm}$ is suitable for prediction on large data. $\texttt{brglm}$ achieves the highest out-of-sample accuracy in the quadratic example, but $\texttt{brglm}$ may not be suitable for big datasets because of relatively expensive computational cost. Meanwhile, the linear model performs well in spite of binary outcome. It shows comparable out-of-sample accuracy with small prediction intervals and the lowest computational cost. 

\section{Discussion}

In this work, we investigate the potential of $\texttt{glmdr}$ for inference and prediction through a comparison of its performance relative to several competing approaches. We found that $\texttt{glmdr}$ yielded statistically optimal inferences and provided predictions that were comparable to other approaches. Overall these results are promising, and suggest that future investigation into applying $\texttt{glmdr}$ to quantitative genetics analyses is warranted. 

\subsection{Statistical benefits and implications of using \texttt{glmdr}}

The logistic model is one of the most common statistical models for analysis of binary traits. Although the classic linear model is an attractive option to use because of its popularity, binary outcome variables violate linear regression modeling assumptions such as homoscedasticity and linearity (i.e. Gauss-Markov assumptions) as well as normality. Therefore, even though results from Section \ref{sec:inference} and \ref{sec:prediction} show that the performance of linear model is comparable to the logistic models, we cannot fully utilize asymptotic properties of linear model and make a proper inference such as significance tests for coefficients. Furthermore, linear regression modeling can yield nonsensical estimates or predictions for a conditional success probability. In some settings confidence intervals can be strictly outside of the $[0,1]$ interval. Such a setting is equivalent to stating that uncertainty exists (i.e., the confidence can take on a range of values) while simultaneously stating that there is no uncertainty (i.e., all values of the confidence interval outside of $[0, 1]$ are forced to take on values of exactly $0$ or $1$ due to the basic definition of a probability).

Within the binary outcome modeling framework we found that $\texttt{glmdr}$ to be the most preferable technique for accounting for data separation. This is based on its overall performance across inference and prediction. A strength of $\texttt{glmdr}$ is that it provides the sharpest inferences and the most natural remedy for data separation within the logistic regression modeling framework. Meanwhile, other methods solve the separation problem by switching the modeling paradigm. For example, $\texttt{bayesglm}$ adopts a Bayesian approach which scales the data first and then places Cauchy distribution as a prior distribution on the coefficients, and $\texttt{brglm}$ modifies the score function to produce finite coefficients. As a result, not only are both models' results in inference not the best, but it is also hard to see their outputs as a natural remedy for data separation. Our novel prediction method for $\texttt{glmdr}$ is comparable with \texttt{bayesglm} and \texttt{brglm} in terms of LOOCV prediction error. However, it may be expensive with a large number of observations and a large number of predictors. 

\subsection{Implications of these results for quantitative genetics}

One of the most notable results was that the linear regression model performed comparably to the other models  for both inference and prediction. Thus, it might be tempting to conclude that a linear regression model is sufficient for the purposes of inference and prediction when applied to the quantitative genetics analysis of binary traits. However, the use of a linear regression model on a binary trait is fundamentally flawed for many reasons \citep{agresti_categorical_2013}. As previously alluded to, the most apparent is that it is possible to obtain predicted values of $\pi(x) = \text{Pr}(Y = 1 | X = x) > 1$ or $\pi(x) < 0$. Such a result contradicts the basic definition of a probability, and thus renders the ensuing statistical analysis useless.  

We have clearly demonstrated the usefulness of \texttt{glmdr} for inferences, in that the benefits suggested by statistical theory translates into optimal precision of estimates (Table 1). However, one important shift that will need to occur for quantitative genetics analyses like genome-wide association studies (GWAS) is that inferences will need to be made in terms of one-sided confidence intervals for $\pi(x)$. The use of such confidence intervals will need to replace the traditional practice of using coefficient estimates in the model to quantify additive, dominance, and epistatic genetic signals. In the presence of complete and quasi-complete separation, slope estimates will be `at infinity' and will therefore not be of practical use. However, one can still perform likelihood ratio tests on nested models under such circumstances to infer the significance of genetic signals included in the model as explanatory variables. This nested model hypothesis testing strategy can work even when separation exists in the null model \citep[Section 3.15]{Geyer2009}. Therefore, the presence of additive, dominant, and epistatic genetic effects can still be inferred without looking at coefficient estimates.

To make \texttt{glmdr} amenable for use in GWAS, future research  should investigate the incorporation of random effects so that it can be used in diversity panels where subpopulation structure and differential familial relatedness are present. Essentially, this would lead to a model analogous to the unified mixed linear model \citep{yu2006unified}, which will account for false positive marker-trait in the presence of these two sources of variability. The \texttt{glmdr} method can potentially be used to make predictions when data separation exists within the fixed effects of a mixed effects model. The main challenge is in the development of a fast likelihood maximizing routine which can compute the optimizations in displayed equation 6 in Section 4.3 of \cite{Eck2021}. The main difficulty is that the data separation problem appears as the integrand in the integration of the random effects terms. This difficulty is substantially mitigated in the mixed effects model when data separation appears in a binary predictor since this integrand is monotonic in $\beta$. 

One of the most exciting developments in this study was the development of a novel method that uses \texttt{glmdr} to make predictions. We used this method to show that \texttt{glmdr} is at least as competitive as other approaches for prediction. To expand upon the findings of this study and assess the potential of \texttt{glmdr} for genomic prediction, we will need to to explore the issues of separation for $p \gg n$ situations, as such a situation will be ubiquitous for genomic prediction in the foreseeable future. The next logical step for investigating \texttt{glmdr} in the $p \gg n$ setting would be to use a penalized likelihood based approach to search for a logistic regression model with $k < n < p$ relevant terms where separation is still present. There are two ways forward within this avenue. One is to use a model selection criteria that penalizes model complexity (AIC, BIC, AICc, \citep{konishi2008information}). The biggest challenge in this front is the development of a fast searching algorithm that preserves the ``correct'' $k$ variables. The other way is to use penalties on the size of $\beta$, such as LASSO \citep{tibshirani1996regression} and ridge regression \citep{hoerl1970ridge}, to screen for slope estimates that are 'at infinity'.

\subsection{Conclusions}
The ever-increasing amount of available high-throughput genotypic and phenotypic crop data is making it possible to both study and utilize genotype-to-phenotype relationships. To ensure that important biological conclusions are based on statistically rigorous analyses, it is critical that the most appropriate methods are used. Because we demonstrated that one of the most appropriate methods, maximum likelihood estimation for handling separation in binary trait data \citep{Geyer2009, Eck2021} is capable of adequate inference and prediction, we conclude that $\texttt{glmdr}$ \citep{glmdr_package} is appropriate for quantitative genetics analyses. However, further research should be conducted to further refine $\texttt{glmdr}$ to handle the most challenging issues typically observed in genotype-to-phenotype analyses.

\section*{Acknowledgements}

We are grateful to Ioannis Kosmidis for feedback on the \texttt{brglm} method and helpful comments, and David Firth for helpful comments.

\bibliographystyle{abbrvnat}
\bibliography{main}

\end{document}